# CHALLENGES TOINTEGRATION OF INFORMATION TECHNOLOGY IN PHYSICAL INFRASTRUCTURE DEVELOPMENT PROCESSES AT THE LOCAL GOVERNMENT LEVEL

**Edward Ayebeng Botchway[1], Ezer YeboahBoateng[2], Titus Ebenezer Kwofie[3]**

[1]Department of Architecture, Kwame Nkrumah University of Science and Technology, Kumasi, Ghana.

[2]Ghana Technology University College, Ghana.

[3] Department of Architecture, Kwame Nkrumah University of Science and Technology, Kumasi, Ghana.

## ABSTRACT

*Ghana's Decentralization Policy has made significant contribution in infrastructural implementation and delivery through Metropolitan, Municipal and District Assemblies as an alternative development strategy to ensure the implementation of the overall national agenda. To date, many systems and strategies have been implemented towards improving development of infrastructure at the local government level. Given that initiation and implementation of infrastructural development at the local level involves the multi-institutional participation of various stakeholders in local governance for effective monitoring and development, several governance forms including e-governance have been initiated towards improvement and effective management of the assemblies. E-Governance through the use of relevant Information and Communication Technology (ICT) has been implemented through the Government Policy on Information and Communications Technology for Accelerated Development (ICT4AD) as a catalyst to facilitate the cross-sectoral participation in the implementation of infrastructural development. Unfortunately, to date, the ICT4AD does not appear to be successfully on course having been in existence for over a decade. Similarly, the development and monitoring of the physical infrastructure at the local level continue to suffer several setbacks and it is currently clearly evident that a refined system of best practices must be put in place to simplify and harmonize the entire development process. This paper is aimed at the assessment of the challenges of e-governance in infrastructural development at the local government level in Ghana. Using a mixed approach with purposive sampling, data gathered from the MMDAs in Ashanti Region suggest that the entire programme has been very slow; bedeviled with lack of funding and lack of commitment on the part of government amongst others. It further revealed that the performance and implementation of the ICT4AD Policy requires an overhaul to achieve the set objectives. The findings offer a basis for formulating frameworks for an effective decision support system in infrastructural delivery and governance at the local government level.*
**Keywords:** Decentralization policy, MMDAs, E-governance, Infrastructural development, ICT4AD.

## 1. INTRODUCTION
The digital age has brought with it a new paradigm of information communication and data processing which has become critical to the success of organizations across the globe. As the quantum of information needed increases the rate of processing and disseminating of information has improved. The use of Information Technology (IT) has become the most appropriate way to process information available swiftly and to employ decisions adduced from the analysis of management information and data with precision. Across the globe organizations are investing heavily in the management of their information systems to remain competitive and efficient. These IT programmes, tools and related infrastructure also makes information sharing amongst all stakeholders in organizations smooth and provides an excellent feedback mechanism via the same platform. Nickerson(2000) posits that an Information System (IS) is not only limited to the various components that make up the infrastructure but includes the people who run these systems as well as the various processes that they engage in to achieve results. In effect all these components come together to define an Information System (IS).

Any Information System (IS) could then be defined as a set of Information Technology (IT) tools that could be used for collecting, storing, processing, sharing and displaying information. Some classic examples of Information Systems employed by organizations are Customer Relationship Management (CRM) Systems, Enterprise Resource Planning (ERP) Systems, Supply Chain Management System (SCM), Product Data Management (PDM) Systems, Project Follow-up (PFU) Systems and others that provide efficient and effective ways of making organizational data and information easily accessible for decision making.

Nickerson (2001) believes that organizations will benefit immensely not only by procuring and deploying these systems but also adapting these systems to the organizational processes.

The adoption and deployment of Information and Communication Technologies (ICT) in the civic, public and commercial sector has increased the knowledge base of society and made ICT more familiar to the general working populace (Ebrahim & Irani, 2005).

In the 21$^{st}$ century governments have turned over to the use of ICT to carry out government business. E-government which is simply the use of ICT by government for the sole purposes of governance has become an efficient way of improving government operations and the delivery of services to the public sector. Heeks (2001) and McClure (2000) have indicated that through web-portal, citizens and business entities could complete transactions with government agencies and departments without necessarily walking to the specific physical location of the latter. This according to Heeks (2000) ultimately improves quality of their services, reduces cost and makes it convenient for both government and businesses to interact electronically and may mark the departure from the brick and mortar system of doing government business.

According to Ebrahim&Irani (2005) an e-government strategy is a crucial and critical success factor in modernizing the public sector, and this could be done by identifying and developing the organizational structure of the various departments and finding innovative ways of providing interactions between businesses, citizens and other government agencies and departments. This is expected to reduce cost and bureaucracy in organizational business processes. Government business processes within other government agencies, external private businesses as well as citizens who do business with government could benefit from the deployment of e-government and this could be done by providing a wide variety of specified information platform to citizens and businesses via an extranet, intranet or the internet.

Ebrahim&Irani (2005) further states that e-government can be deployed to develop the strategic connections between public sector organizations and their departments, and provide secured communication between the different government levels of operation (e.g. central, city, and local). Governments can also transfer funds electronically to other governmental agencies or provide information to public employees through an intranet or internet.

In 2003, The Government of Ghana put together a policy statement on ICT for Accelerated Development (ICT4AD). This policy document sought to integrate ICT into every fiber of society in Ghana and particularly into government services machinery for accelerated development and to ensure that by the Year 2022 it would have rolled out the final of the five phase comprehensive plan. The policy statement (ICT4AD, 2003, p. 41) gave a clear indication of government's intention to embark on e-government by facilitating government administration and service delivery electronically across board for all Public Sector Organizations(PSO) , Regional Coordinating Councils( RCC), District Councils, Public Enterprise Commission(PEC), Trade and Labour Unions and Associations and also at the local government levels which is currently manned by the Metropolitan, Municipal, District Assembly (MMDAs). The main objective was to promote effectiveness and efficiency in service delivery across MMDAs and other governmental departments/agencies and ultimately bring government closer to the people.

A United Nations Public Administration Report(2012) issued after UN e-government Survey (2012) indicated that Ghana has not done well as well as Sub-Saharan Africa which has an e-government Index of about 0.3 as against a European index of 0.7.This gave an indication that Ghana is still at the Emerging Phase of e-government. A critical look at the situation on the ground indicated that not much has been done in this area when it comes to the implementation of this policy. This is not surprising since there is no proper documentation regarding progress of same. It is believed from the documentation adduced that government has not allocated enough resources to see to the implementation of this policy.

It is important to note however that every establishment including that of government is confronted with the challenge of harnessing all the vital information available to its management for purposes of effective decision making and to gain competitive advantage. Salaazar et al (2007) believe that typical organizations have 90% of all their records and proceedings done manually and on paper and most of the ICT and Information Systems in operation are not fully integrated.

In the case of public sector organizations in developing countries like Ghana particularly the Metropolitan, Municipal and District Assemblies the situation is worse. The request of a simple document will demand that you move from one office to another, one department to another department. Huge books of reference will have to be opened and cross-checked to find your details. Paper files are still difficult to trace and in a singular transaction one will have to contact several departments for the transaction to be completed. Even for relatively simple matters like building permit acquisition, one must be prepared to physically go to the Metropolitan Town and Country Planning Department, Metropolitan Works Department, Metropolitan Physical Planning Department, Metropolitan Urban Roads Department and a number of different agencies who may be involved in this single transaction.

The study therefore seeks to find out the state and benefits of e-government/ Information Management Systems in the Metropolitan, Municipal and District Assemblies in Ashanti Region and to find ways of improving the performance in service delivery of these MMDAs using Information Technology. The findings of this study will be a vital piece of information to MMDAs to assess their IT/ISM readiness and suggest ways of improving the infrastructure and processes. It will also afford MMDAs the opportunity to examine their current location in their journey towards a full local e-government platform and how to deploy e-government to their communities for effective development of their infrastructure. The study could be beneficial to other Metropolitan, Municipal and District Assemblies in other regions in Ghana who are embarking on e-government aggressively.

## 2 METHODS

The study was conducted in Ashanti Region which currently has thirty (30) districts made up of one (1) Metropolitan Assembly, eight (8) Municipal Assemblies and twenty one (21) District assemblies and one (1) Regional Coordinating Council. Ashanti Region has a population of over three million people and has second largest number of districts. The region has a total land surface of 24,389 km sq. which is 10.2 percent of the total land area of the country, Ghana.

The research employed the quantitative approach as well as the qualitative approach to data gathering to ascertain the quantum and state of IT infrastructure in place at the various assemblies.

The sample size determination for the assemblies was based on a formula provided by a generic sample size calculation formula which takes into consideration the required sample size, the population size for the entire target population, the population proportion, and the degrees of accuracy and freedom that is being expressed as a proportion.

Different data collection instruments such as questionnaires, interviews, backed by a checklist were used. Both primary and secondary data were also elicited in this study. In gathering the primary data, questionnaires, semi-structured interviews, observation, as well as some focused group discussions was undertaken. The sampled number of assemblies targeted was a little over fifty (50) per cent of all MMDAs in the region. The websites of all the assemblies were reviewed to ascertain the extent of the state of e-government preparedness and also assess their position in the journey towards achieving full e-government.

Questionnaires were sent to hundred (100) personnel in the assemblies to solicit their views on Information Technology/Management Information Systems and e-governance issues out of which eighty (80) people responded to the questionnaire administered. The Likert scale was employed in the designing of the questionnaire.

The secondary source used was through the review of text books, articles in newspapers, popular magazines, journals and publications, research papers, websites, policy framework, and reports on Management Information Systems and e-governance.

The analysis for the collected data was done using Microsoft Excel and SPSS. Descriptive statistics was the mode used for analysis. The statistical analysis was corroborated by the interviews conducted and observations by the researcher. The researcher employed research best practices such as ethical considerations to collect data from the study population. Having gotten a response rate of about 80% from the sampled population, it could be concluded that the conclusions drawn from the study are reliable and valid.

**Table 1:** CATEGORIES OF MMDAs: Target population vs. Sampled Population Sizes

| SERIAL NO. | CATEGORIES OF ASSEMBLIES IN ASHANTI REGION | ASSEMBLIES IN ASHANTI REGION | TARGET POPN. | SAMPLED POPN. | NO. OF RESPONDENTS |
|---|---|---|---|---|---|
| 1 | METROPOLITAN | KMA | 1 | 1 | 5 |
| 2 | MUNICIPAL | • Obuasi<br>• Offinso<br>• Mampong<br>• Ejura-Sekyeredumase<br>• Ejisu-Juaben<br>• Bekwai<br>• Asante Akim Central<br>• Asokore Mampong | 8 | 5 | 5 |

| 3 | DISTRICT | • AtwimaKwanwoma<br>• AtwimaMponua<br>• AtwimaNwabiagya<br>• Afigya-Kwabre<br>• Ahafo-Ano South<br>• Ahafo-Ano North<br>• Kwabre-East<br>• Bosomtwe<br>• Asante-Akim South<br>• Asante-Akim North<br>• Bosome Freho<br>• Sekyere Afram Plains<br>• Offinso North<br>• Sekyere Central<br>• Sekyere East | 21 | 10 | 5 |
| --- | --- | --- | --- | --- | --- |
|  |  | • Sekyere South<br>• Kumawu<br>• Adansi North<br>• Adansi South<br>• Amansie Central<br>• Amansie West |  |  |  |
|  | TOTAL |  | 30 | 16 | 80 |

## 3 RESULTS AND DISCUSSION
### 3.1 CURRENT STATE OF INFORMATION TECHNOLOGY/INFORMATION SYSTEM
#### 3.1.1. Kumasi Metropolitan Assembly
All the respondents from KMA confirmed that there are computers in KMA. As to the number of the computers in KMA, the respondents gave varying figures and it is understandable that they have little knowledge as to the exact computers in KMA because KMA is bigger. The numbers given by the IT personnel was used because it is their line of duty. Deducing from the IT personnel KMA has between 20-50 computers. In describing the IT infrastructure in KMA, some of the respondents stated that there are only standalone computers used for word and spreadsheet processing. However the respondent from the IT section indicated that there are some networked computers used for email and MIS. The respondents attested that there is an IT department and this was confirmed by the researcher's observation through his visit. The respondents stated that the IT infrastructure is not outsourced to a third party and the assembly partly maintains some records in digital form even though the bulk of the data is still manually and paper processed. The respondents indicated that KMA does not have its own automated platform for system integration. In disseminating information across various departments, the respondents stated that it is mainly done in verbal form, putting the information on the notice board for all staff to read and by letters delivered manually to all departments. In terms of storage of data, the respondents all indicated that it is by paper and files. However the respondents from the IT section indicated that data is sometimes stored electronically. In disseminating information to the community, the respondents indicated that it is done through radio, information vans, electronically and sometimes pasting of posters.

In the usage of software, the respondents indicated that finance department and general administration uses Microsoft suite.

### 3.1.2. Municipal Assemblies

All the respondents in the 5 municipal assemblies sampled stated that the assemblies have computers. In terms of the number of the computers, the table below shows the quantities.

**Table 2:** Number of computers in Municipal Assemblies

| Serial No. | District | No. of Computers |
|---|---|---|
| 1. | Asante –Akim Central, Konongo | 21-40 |
| 2. | Ejisu-Juabeng | 1-20 |
| 3. | Obuasi | 21-40 |
| 4. | Asokore Mampong | 1-20 |
| 5. | Bekwai | 1-20 |

Source: Field data (2013)

According to respondents from each assembly, Konongo has between 21-40 computers, Ejisu-Juaben has between 1-20 computers, Obuasi has between 21-40 computers, Asokore Mampong and Bekwai has 1-20 computers each. All the respondents from the 5 municipal assemblies stated that the computers are standalone computers for word and excel processing. The researcher through his visit also observed the status of the computers as being just standalone computers. None of the municipal assemblies have an IT department but rather have IT units that coordinates and maintains computer related processes in the assembly.

All the respondents from the 5 assemblies stated that the IT infrastructure is not outsourced to a third party and the assembly maintains no confidential records in digital format. The respondents also indicated that the assemblies do not have its own automated platform for system integration. In disseminating information across various departments, the respondents stated that it is mainly done verbally and the use of notice boards for staff and community information. In terms of storage of data, all the respondents indicated that it is by paper and files. In disseminating information to the community, the respondents from Obuasi, Konongo and Ejisu, Asokore Mampong and Bekwai all indicated that it is done through radio, information vans, sometimes pasting of posters.

### 3.1.3. District Assemblies

All the respondents from the 10 district assemblies confirmed that there are computers in their assemblies. As to the number of the computers in their respective assemblies, all but Atwima Kwawoma and Nwabiagya stated that they have 1-20 computers. Respondents from Atwima Kwawoma and Atwima Nwabiagya indicated that they have between 21-40 computers. In describing the IT infrastructure in their assemblies all the respondents in the 10 district assemblies stated that there only standalone computers used for word and spreadsheet processing. Through the visit made by the researcher, the researcher also observed standalone computers in various offices which were not networked. All the respondents indicated that their respective assemblies do not have neither IT departments nor units but have an individual who helps in resolving IT related issues and this was confirmed by the researcher's observation through his visit. The respondents stated that the IT infrastructure is not outsourced to a third party and the assembly maintained some records in digital formats. The respondents indicated that the assemblies do not have its own automated platform for system integration. In disseminating information across various departments, the respondents stated that it was mainly done verbally and by posting notices on the boards for workers, departments and community. In disseminating some information to the community, the respondents indicated that it is done by information vans, and sometimes pasting of posters. In the use of software, the respondents indicated that finance department and general administration uses Microsoft suite.

### 3.2. CHALLENGES ASSOCIATED WITH THE INTEGRATION OF IT/ISM

### 3.2.1. Financial and Economic Challenges

As indicated by Table 3 below, the cost of MIS was view by respondents as the most challenging with the highest of mean of 4.31. 60 of the respondents considered it as very challenging, 11 also indicated it as an important challenge whiles 4 view it as a minor challenge. Cost for providing MIS Services also had a mean of 4.19 and was the second highest. Cost for governments of meeting laws and regulations relating to MIS had a mean of 3.24 whiles creating special funds for MIS Implementation had a mean of 2.91. Recording the least mean of 2.17 was difficulty in demonstrating the long term cost-benefits of MIS initiatives. 68 of the respondents considered it as just a minor challenge whiles 3 and 4 of the respondents considered it as important and a very important challenge respectively. The average mean was 3.36

Table 3: Financial and Economic Challenge

| SERIAL NO. | VARIABLE | RESPONSE | | | | | Rank |
|---|---|---|---|---|---|---|---|
| | | Not a Challenge | A Minor Challenge | Don't Know | An Important Challenge | A Very Important Challenge | |
| 1. | Cost of developing MIS Services | 0 | 4 | 0 | 11 | 60 | 4.31 |
| 2. | Cost for providing MIS Services | 0 | 3 | 0 | 14 | 58 | 4.19 |
| 3. | Cost for governments of meeting laws and regulations relating to MIS | 0 | 4 | 0 | 31 | 40 | 3.24 |
| 4. | Difficulty in demonstrating the long term cost-benefits of MIS initiatives | 0 | 68 | 0 | 3 | 4 | 2.17 |
| 5. | Creating special Funds for MIS Implementation | 0 | 21 | 11 | 19 | 24 | 2.91 |
| | | | | | | Average Mean | 3.36 |

Source: Field data (2013)

**3.2.2. Access, Skill and Usage Challenge**
Low level of internet/ICT use among citizen groups received the highest and second highest means respectively as very important challenges to the implementation of MIS in the assemblies. The mean for Low level of internet use among citizen groups was 4.57 whiles the mean for ICT skills among Citizens was 4.56. 49 of the respondents stated that Low level of internet use among citizen groups was a very important challenge, 23 saw it as an important challenge. Only three view it as a minor challenge. 52 of the respondents also considered ICT skills among Citizens as a very important challenge whiles 18 saw it as an important challenge. Only 5 regarded it as a minor challenge. Recording the lowest mean was Citizens lack strong motivations to use of electronic services which had a mean of 3.21. The average mean was 3.98

Table 4: Access, Skill and Usage Challenge

| SERIAL NO. | VARIABLE | RESPONSE | | | | | Rank |
|---|---|---|---|---|---|---|---|
| | | Not a Challenge | A Minor Challenge | Don't Know | An Important Challenge | A Very Important Challenge | |
| 1. | Low level of internet use among citizen groups (eg: relating to age, income, literacy, education, etc. | 0 | 3 | 0 | 23 | 49 | 4.57 |
| 2. | ICT skills among Citizens | 0 | 5 | 0 | 18 | 52 | .56 |
| 3. | ICT Skills among Government Officials | 6 | 37 | 0 | 11 | 21 | 3.05 |
| 4. | Public perception of risks to privacy and civil | 1 | 13 | 0 | 27 | 34 | 4.07 |
|    | liberties |   |   |   |   |   |   |
| 5. | Public concerns over potential for online theft and fraud | 2 | 2 | 0 | 28 | 43 | 4.44 |
| 6. | Citizens lack strong motivations to use of electronic services | 15 | 12 | 8 | 22 | 18 | 3.21 |
|    |    |    |    |    |    | Average mean | 3.98 |

Source: Field data (2013)

**3.2.3. Technical and Design Challenge**

Difficulty in using MIS applications had the highest mean of 2.65 followed by Difficulty in using MIS applications with a mean of 2.51. The challenge with the least mean was lack of Interoperability between IT Systems with a mean of 2.17. The average mean for technical and design challenge was 2.36. The means for the individual challenges within the technical challenge was low due to the fact that most of the respondents lack the technical understanding of this challenge as such most of them ticked the "don't know" column resulting in lower means.

**Table 5:** Technical and Design Challenge

| SERIAL NO. | VARIABLE | RESPONSE | | | | | |
|---|---|---|---|---|---|---|---|
|   |   | Not a Challenge | A Minor Challenge | Don't Know | An Important Challenge | A Very Important Challenge | Rank |
| 1. | Variety of languages used in your country/language barrier on systems | 26 | 34 | 0 | 8 | 7 | 2.15 |
| 2. | Making MIS services easily accessible to the visually impaired & other disabilities | 19 | 24 | 16 | 14 | 2 | 2.41 |
| 3. | Difficulty in using MIS applications | 16 | 27 | 17 | 8 | 7 | 2.51 |
| 4. | Lack of secure electronic Authentication and Identification | 15 | 23 | 18 | 11 | 8 | 2.65 |
| 5. | Lack of Standards for electronic Identification in your country | 24 | 21 | 21 | 5 | 4 | 2.25 |
| 6. | Lack of Interoperability between IT Systems | 21 | 23 | 29 | 1 | 1 | 2.17 |
|   |   |   |   |   |   | Average Mean | 2.36 |

Source: Field data (2013)

**3.2.4 Regulatory Challenge**

As indicated by Table 6, enacting and implementing cyber crime related laws according the respondents was ranked the highest with a mean of 4.24 followed by absent of collection, sharing, use and control of personal data and data protection law had a mean of 3.37. Effective application of Authentication/ Identification Laws and Regulations had a mean of 3.36 whiles Legal Concerns with Public Private Partnership for Infrastructure had a mean of 3.31. The challenge with the least mean is employment law that constrains IT enabled structuring of jobs with a mean of 2.6. The average mean was 3.20.

**Table 6:** Regulatory Challenge

| SERIAL NO. | VARIABLE | RESPONSE | | | | | |
|---|---|---|---|---|---|---|---|
| | | Not a Challenge | A Minor Challenge | Don't Know | An Important Challenge | A Very Important Challenge | Rank |
| 1. | Effective application of Authentication/ Identification Laws and Regulations | 11 | 9 | 18 | 16 | 21 | 3.36 |
| 2. | Inadequate policies on freedom of Information | 16 | 19 | 23 | 10 | 7 | 2.64 |
| 3. | Legal Concerns with Public Private Partnership for Infrastructure | 9 | 7 | 32 | 6 | 21 | 3.31 |
| 4. | Employment law that constrain IT Enabled structuring of jobs | 6 | 27 | 33 | 9 | 0 | 2.60 |
| 5. | Absent of collection, sharing, use and control of personal data and data protection Law | 9 | 7 | 24 | 17 | 18 | 3.37 |
| 6. | Lack of general rights for citizens to communicate electronically with public authorities | 11 | 6 | 38 | 8 | 12 | 3.05 |
| 7. | Copyright Protection | 13 | 9 | 29 | 11 | 13 | 3.03 |
| 8. | Heightened risks of liability | 16 | 11 | 15 | 11 | 22 | 3.16 |
| 9. | Enacting and implementing cyber-crime related laws | 0 | 12 | 0 | 21 | 42 | 4.24 |
| | | | | | | Average Mean | 3.20 |

Source: Field data (2013)

**3.2.5 Organizational and Administrative Challenge**
Table 7, indicates that resistance to change by government officials recorded the highest mean of 4.49. 55 of the respondents regarded it as a very important challenge and a further 12 considered it as an important challenge. 6 saw it as a minor challenge whiles 2 view it as not a challenge. Coordination across central, regional and Local Government recorded the second highest mean of 4.27. The challenge with the least mean was lack of policy support for MIS project with a mean of 4.11. The average mean is 4.26

**Table 7:** Organizational and Administrative Challenge

| SERIAL NO. | VARIABLE | RESPONSE | | | | | |
|---|---|---|---|---|---|---|---|
| | | Not a Challenge | A Minor Challenge | Don't Know | An Important Challenge | A Very Important Challenge | Rank |
| 1. | Resistance to change by government officials | 2 | 6 | 0 | 12 | 55 | 4.49 |
| 2. | Coordination across central, regional and Local Government | 1 | 3 | 12 | 18 | 41 | 4.27 |
| 3. | Coordination Between Public Administration, Citizen and other Actors | 7 | 3 | 6 | 13 | 46 | 4.17 |
| 4. | Lack of policy support for MIS project | 6 | 5 | 7 | 14 | 43 | 4.11 |
| | | | | | | Average Mean | 4.26 |

Source: Field data (2013)

### 3.2.6 Summary of Challenges

From Table 8 below, organizational and administrative challenge had the highest average mean of 4.26 and constitutes 25% of the total challenges put together. It is followed by the challenge of Access, skill and usage with an average mean of 3.98 and also constituted 23% of the challenges put together. Financial and economic challenges were third with an average mean of 3.36 and constituted 20% of the challenges put together. Next was regulatory challenge with an average mean of 3.20 and constituted 19% of the challenges. The least challenge was Technical and Design which had an average mean of 2.36 and constituted 14% of the challenges.

**Table 8:** Summary of challenges

| Age | Av. Mean | Percent | Valid Percent | Cumulative Percent |
|---|---|---|---|---|
| Financial and Economic | 3.36 | 20 | 20 | 20 |
| Access, skill and usage | 3.98 | 23 | 23 | 43 |
| Technical and Design | 2.40 | 14 | 14 | 57 |
| Regulatory | 3.20 | 19 | 19 | 75 |
| Organizational and Administrative | 4.26 | 25 | 25 | 100.0 |
| Total | 17.20 | 100.0 | 100.0 | |

Source: Field data (2013)

### 3.3 BENEFITS OF MIS AND E-GOVERNMENT

All the respondents indicated that their respective assemblies do not engage in extensive electronic business but perceived MIS/E-government as beneficial.

### 3.3.1. Government to Government (G2G)

As indicated by Table 9 and shown in Fig.1, the respondents perceived improves accessibility to data/information and improves information for growth of local economy and better external relations as very beneficial with 3.8 and 3.76 as the means. In terms of convenience, 49 of the respondents perceived it as very beneficial whiles 24 perceived it as beneficial. A further 2 viewed it as not beneficial. The least beneficial item was improvement in local democracy with the lowest mean 3.31. Concerning improvement in local democracy 45 perceived it to be very beneficial, 15 perceived it as beneficial, 8 were not sure of its benefits and 7 viewed it as not beneficial.

**Table 9:** Government to Government (G2G)

| SERIAL NO. | VARIABLE | RESPONSE | | | | |
|---|---|---|---|---|---|---|
| | Perceived benefits | Not Beneficial | Not Sure | Beneficial | Very Beneficial | Rank |
| 1. | Convenience | 2 | 0 | 24 | 49 | 3.60 |
| 2. | Cost effectiveness | 3 | 2 | 19 | 51 | 3.57 |

| 3. | Improves Accessibility to data/information | 0 | 0 | 15 | 60 | 3.80 |
| 4. | Increases internal efficiency | 4 | 5 | 13 | 53 | 3.53 |
| 5. | Improves local democracy | 7 | 8 | 15 | 45 | 3.31 |
| 6. | Improves information for growth of local economy and better external relations | 1 | 2 | 11 | 61 | 3.76 |
| | | | | | Average mean | 3.6 |

Source: Field data (2013)

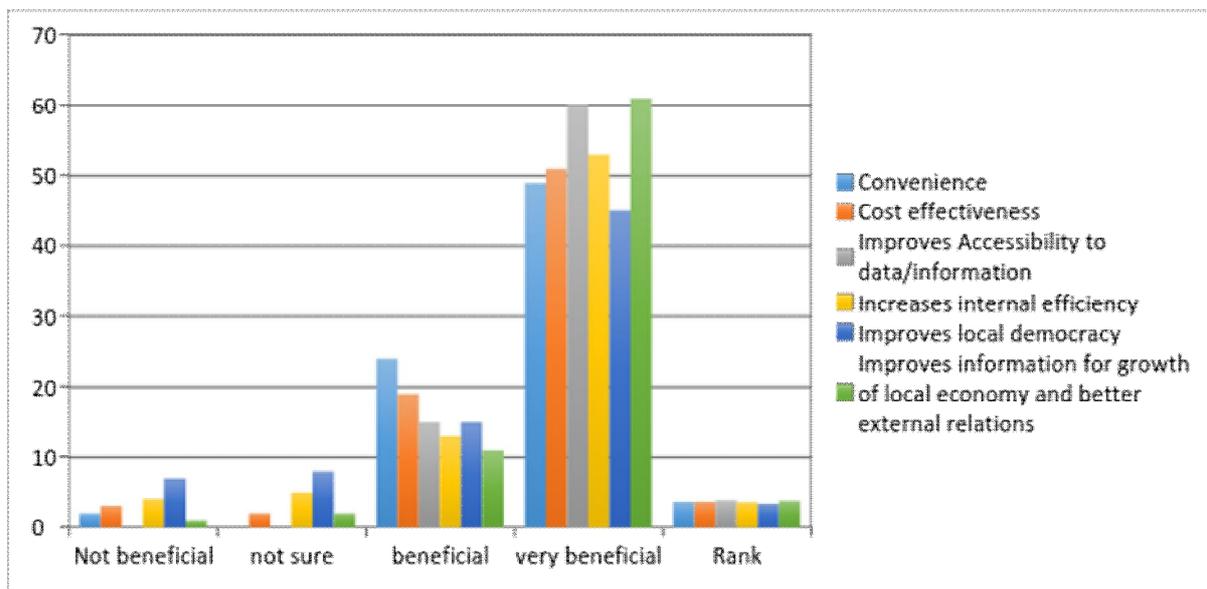

**Figure 1:** Government to Government, Source: Field data (2013)

### 3.3.2. Government to Citizen (G2C)
Concerning G2C, the perceived benefits that recorded the highest mean was improvement in information for growth of local economy and better external relations with a mean of 3.95. 71of the respondents perceived it to be very beneficial and a further four perceived it to be beneficial with none thinking otherwise. Increases internal efficiency recorded the least mean of 3.29.

**Table 10:** Government to Citizen (G2C)

| SERIAL NO. | VARIABLE | RESPONSE | | | | |
|---|---|---|---|---|---|---|
| | | Not Beneficial | Not Sure | Beneficial | Very Beneficial | Rank |
| 1. | Convenience | 0 | 3 | 13 | 59 | 3.75 |
| 2. | Cost effectiveness | 1 | 2 | 17 | 55 | 3.68 |
| 3. | Improves Accessibility to data/information | 0 | 1 | 9 | 65 | 3.85 |
| 4. | Increases internal efficiency | 2 | 0 | 47 | 26 | 3.29 |
| 5. | Improves local democracy | 1 | 1 | 37 | 36 | 3.44 |
| 6. | Improves information for growth of local economy and better external relations | 0 | 0 | 4 | 71 | 3.95 |
| | | | | | Average mean | 3.66 |

Source: Field data (2013)

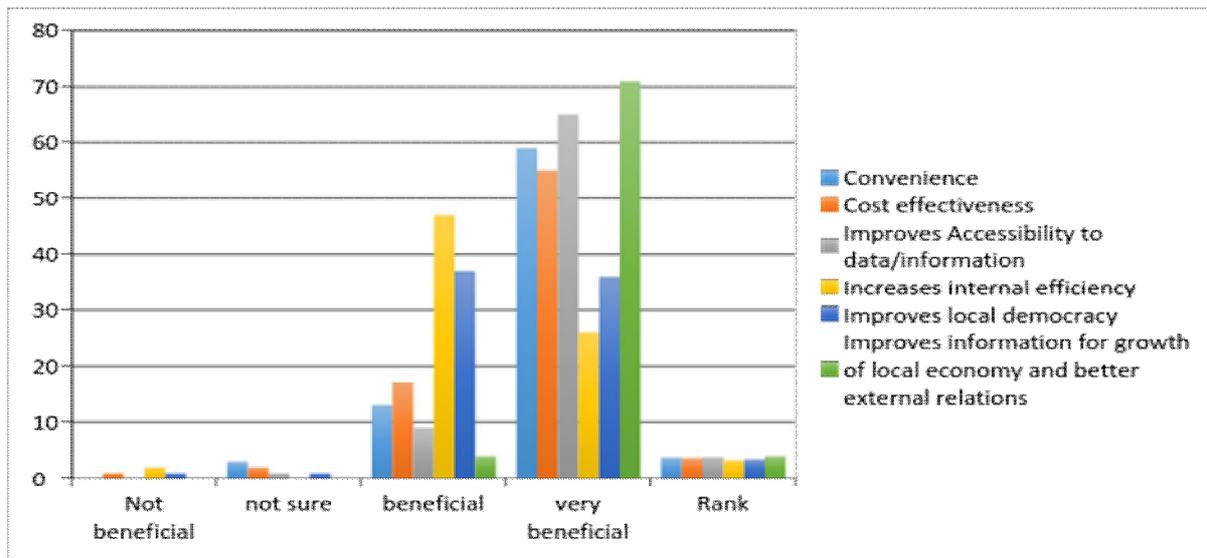

**Figure 2:** Government to Citizen, Source: Field data (2013)

### 3.3.3. Government to Business (G2B)
The Benefit which was perceived by respondents as being very beneficial is improvement information for growth of local economy and better external relations with a mean of 3.52. Improves Accessibility to data/information had a mean of 3.36 whiles convenience had a mean of 3.42. The benefit with the least mean was cost effectiveness with a mean of 3.31.

**Table 11:** Government to Business (G2B)

| SERIAL NO. | VARIABLE | RESPONSE | | | | |
|---|---|---|---|---|---|---|
| | | Not Beneficial | Not Sure | Beneficial | Very Beneficial | Rank |
| 1. | Convenience | 1 | 1 | 38 | 34 | 3.42 |
| 2. | Cost effectiveness | 1 | 4 | 41 | 29 | 3.31 |
| 3. | Improves Accessibility to data/information | 3 | 4 | 31 | 37 | 3.36 |
| 4. | Increases internal efficiency | 0 | 1 | 26 | 48 | 3.63 |
| 5. | Improves local democracy | 2 | 4 | 26 | 43 | 3.47 |
| 6. | Improves information for growth of local economy and better external relations | 5 | 2 | 17 | 51 | 3.52 |
| | | | | | Average Mean | 3.45 |

Source: Field data (2013)

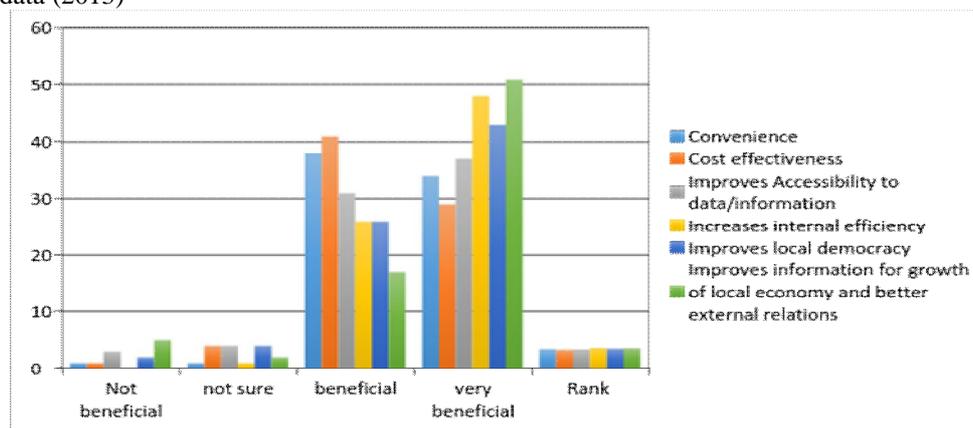

**Figure 3:** Government to Business, Source: Field data (2013)

### 3.3.4. Government to Employee (G2E)
In terms of G2E, increase in internal efficiency recorded the highest mean of 3.65 as such regarded the most beneficial factor. The beneficial factor with the lowest mean was cost effectiveness with a mean of 3.21. Improves information for growth of local economy and better external relations had a mean of 3.44 whiles Improvement in accessibility to data/information recorded a mean of 3.55.

**Table 12:** Government to Employee (G2E)

| SERIAL NO. | VARIABLE | RESPONSE | | | | |
|---|---|---|---|---|---|---|
| | | Not Beneficial | Not Sure | Beneficial | Very Beneficial | Mean |
| 1. | Convenience | 3 | 1 | 29 | 42 | 3.47 |
| 2. | Cost effectiveness | 1 | 6 | 44 | 24 | 3.21 |
| 3. | Improves Accessibility to data/information | 3 | 4 | 17 | 51 | 3.55 |
| 4. | Increases internal efficiency | 1 | 1 | 21 | 52 | 3.65 |
| 5. | Improves local democracy | 2 | 0 | 34 | 39 | 3.47 |
| 6. | Improves information for growth of local economy and better external relations | 3 | 3 | 27 | 42 | 3.44 |
| | | | | | Average mean | 3.46 |

Source: Field data (2013)

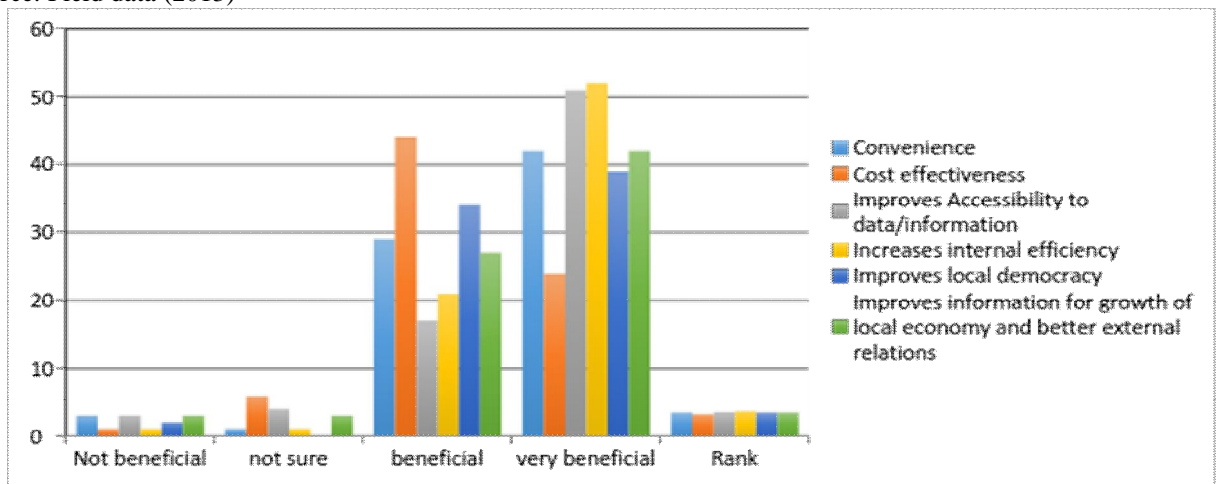

**Figure 4:** Government to Employee, Source: Field data (2013)

### 3.3.5 Summary of Ranking of benefits
From the Table 13 below, it indicates that the perceived benefit of improvement in information management systems for growth of local economy and better external relations recorded the highest overall mean of 14.67. Thus the respondents perceived that MIS and e–government will enable the assemblies to improve information management for growth of local economy and better external relations. Next in the ranking is improvement accessibility to data/information with a total mean of 14.56. It is followed by convenience of a mean of 14.23. Next is increase in internal efficiency which recorded a total mean of 14.11. It is followed by cost effectiveness with a total mean of 13.77.Improvement in local democracy recorded a mean of 13.68.

**Table 13:** Summary of ranking of benefits

| SERIAL NO. | VARIABLE | RESPONSE | | | | |
|---|---|---|---|---|---|---|
| | Benefits | G2G | G2C | G2B | G2E | Total Mean |
| 1. | Improves information for growth of local economy and better external relations | 3.76 | 3.95 | 3.52 | 3.44 | 14.67 |
| 2. | Improves Accessibility to data/information | 3.80 | 3.85 | 3.36 | 3.55 | 14.56 |
| 3. | Convenience | 3.60 | 3.75 | 3.42 | 3.47 | 14.23 |
| 4. | Increases internal efficiency | 3.53 | 3.29 | 3.63 | 3.65 | 14.11 |
| 5. | Cost effectiveness | 3.57 | 3.68 | 3.31 | 3.21 | 13.77 |
| 6. | Improves local democracy | 3.31 | 3.44 | 3.47 | 3.47 | 13.68 |
| | | | | | Average mean | 14.17 |

Source: Field data (2013)

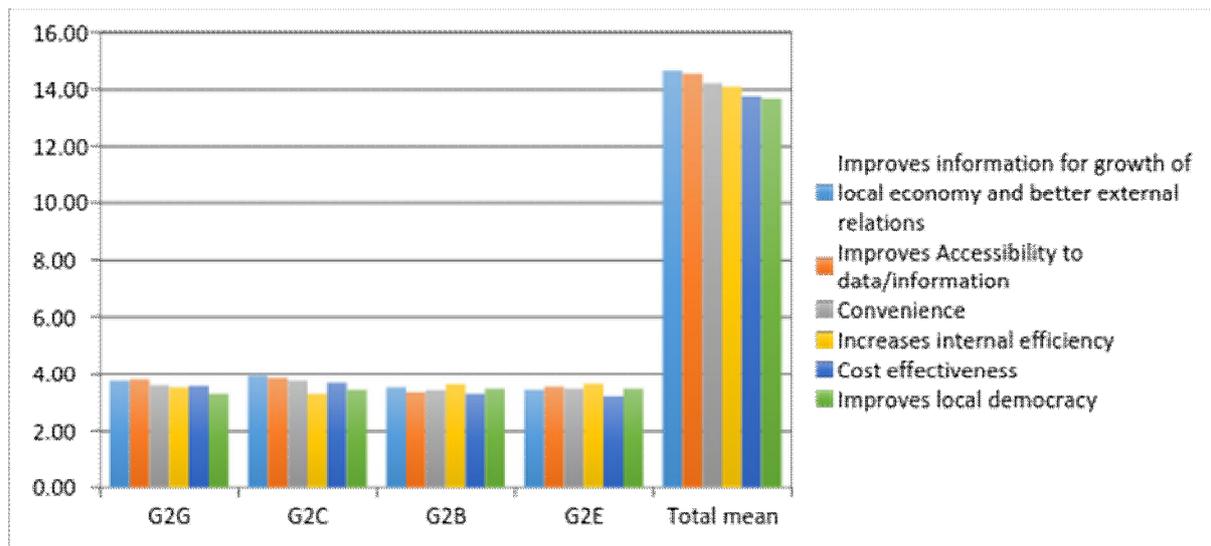

**Figure 5:** Summary of ranking of benefits, Source: Field data (2013)

## 4. CONCLUSION

From the findings, it can be concluded that the current state of the IT infrastructure and human capital at the metropolitan/municipal/district assemblies is not adequate for any effective management of the information system and to embark on e-governance for any meaningful development including infrastructure. An effective IT/ISM and e-governance approach requires modern state of the art facilities to run. There are many benefits that IT/ISM and e-governance brings to bear on the state and its citizens. It will make life easier for citizens to communicate and receive services without having to physically go to the Assembly's office to request for any service. There are many challenges that need to be overcome by the state, assemblies and the citizens to be able to fully deploy the e-governance structure. Currently it can be argued out categorically that the IT/ISM/e-governance is at the Emerging Stages of its development and a lot more will be required to take it to the next phase which is the Enhanced Phase and then subsequently push it further to the Interactive Phase.

**ACKNOWLEDGEMENT**

We acknowledge the contribution of Research Initiatives in Innovation and Technology and Architecture (RIITA) and Projekt David Foundation to ensure that this paper gets published. We appreciate the input of Edward Ayebeng Botchway for using part of his thesis data to corroborate this research. We also thank Mrs. Emelia Botchway for helping us gain access to the assemblies and finally to all the MMDAs sampled. We say a big thank you to the project assistant Rebecca Agyemang-Yeboah.

## REFERENCES

[1]. Ebrahim, Z., & Irani, Z. (2005). E-government adoption: architecture and barriers. Business Process Management Journal, 11(5), 589-611
[2]. Heeks, R. (2001), Understanding E-Governance for Development, Institute for Development Policy and Management, Manchester.
[3]. Heeks, R. (2003). Information Systems and Developing Countries: Failure, Success, and LocalImprovisations. Routledge Taylor, 18(2), 101-112.
[4]. McClure, D. (2000), "Electronic government: federal initiatives are evolving rapidly but they face significant challenges", Accounting and Information Management Division, available at: www.gao.gov/new.items/a200179t.pdf

[5]. Nickerson, R.C. 2000, Business and information systems, Prentice Hall.
[6]. Salazar, A.J. & Sawyer, S. (2007), "Introduction" in Handbook of information technology inorganizations and electronic markets, eds. A.J. Salazar & S. Sawyer, World Scientific,pp. 1-12.
[7]. The Ghana ICT for Accelerated Development (ICT4AD) Policy (2003)
[8]. UN e-government Survey (2012)
[9]. United Nations Public Administration Report (2012)